# Pulse-shape discrimination and energy resolution of a liquid-argon scintillator with xenon doping

Christopher G. Wahl,[a,b] Ethan P. Bernard,[a] W. Hugh Lippincott,[a,c] James A. Nikkel,[a,d] Yunchang Shin,[a,e] and Daniel N. McKinsey[a*]

[a] *Yale University,*
   *New Haven, CT, USA*
   E-mail: daniel.mckinsey@yale.edu

[b] *Now at H3D, Inc.*
   *Ann Arbor, MI, USA*

[c] *Now at Fermilab,*
   *Batavia, IL, USA*

[d] *Now at Royal Holloway, University of London,*
   *Egham, Surrey, UK*

[e] *Now at TRIUMF,*
   *Vancouver, BC, Canada*

ABSTRACT: Liquid-argon scintillation detectors are used in fundamental physics experiments and are being considered for security applications. Previous studies have suggested that the addition of small amounts of xenon dopant improves performance in light or signal yield, energy resolution, and particle discrimination. In this study, we investigate the detector response for xenon dopant concentrations from $9 \pm 5$ ppm to $1100 \pm 500$ ppm xenon (by weight) in 6 steps. The 3.14-liter detector uses tetraphenyl butadiene (TPB) wavelength shifter with dual photomultiplier tubes and is operated in single-phase mode. Gamma-ray-interaction signal yield of $4.0 \pm 0.1$ photoelectrons/keV improved to $5.0 \pm 0.1$ photoelectrons/keV with dopant. Energy resolution at 662 keV improved from $(4.4 \pm 0.2)\%$ ($\sigma$) to $(3.5 \pm 0.2)\%$ ($\sigma$) with dopant. Pulse-shape discrimination performance degraded greatly at the first addition of dopant, slightly improved with additional additions, then rapidly improved near the end of our dopant range, with performance becoming slightly better than pure argon at the highest tested dopant concentration. Some evidence of reduced neutron scintillation efficiency with increasing dopant concentration was observed. Finally, the waveform shape outside the TPB region is discussed, suggesting that the contribution to the waveform from xenon-produced light is primarily in the last portion of the slow component.

---

[*] Corresponding author.

**Contents**

## 1. Introduction

Liquid argon (LAr) has a number of advantageous properties for use as an active material for detection of ionizing radiation. It is extremely inexpensive – only a few dollars per kilogram. Since it is a fluid, it can easily be scaled to large volumes for higher efficiency. Interactions in liquid argon produce large amounts of scintillation, comparable to the light output of NaI(Tl). Therefore, energy resolution can be as low as a few percent FWHM [1]. Unlike NaI(Tl), which has a slow 230-ns decay time, argon has a fast component with a 7-ns decay time, followed by a slower component with 1.6-μs decay time [2]. The fast component allows precise coincidence measurements and potentially high-count-rate measurements. Additionally, charge created in the interaction can be drifted and measured by application of an electric field. The combination of charge and light can be used to improve energy resolution or to identify the interacting particle (often called discrimination) [3, 4]. Using light-collection patterns or charge-collection locations, precise position estimation is also possible. Table I lists some properties of liquid argon.

     Even when operated as a simple scintillator without any charge collection, liquid argon provides pulse shape discrimination (PSD) between electronic recoils produced by gamma-ray or beta-particle interactions and nuclear recoils produced by neutrons (or dark matter) [5]. The PSD arises because interactions of ionizing radiation in argon produce both ions and excited argon atoms. The excited atoms often bond with ground state atoms to form metastable molecules known as excimers that then decay, emitting scintillation. The ions can recombine with electrons to form excited atoms, in turn producing excimers and then scintillation light.

TABLE I. PROPERTIES OF ARGON AND XENON (DATA FROM [4], [6], AND [7])

| Property | Argon | Xenon |
|---|---|---|
| Fast decay time (ns) | 7 | 4.3 |
| Slow decay time (ns) | 1600 | 22 |
| Light yield (photons/keV) | 40 | 42 |
| Wavelength (nm) | 128 | 175 |
| Density at boiling temperature at 1 atm (g/cm$^3$) | 1.40 | 2.94 |
| Cost (US$/kg) | ~2 | ~2000 |

The excimers can exist in two states, known as singlet and triplet. It is thought that the ratio of atomic excitations to ionizations is much greater for nuclear recoils than for electron recoils [8, 9]. Direct atomic excitation primarily results in singlet-state excimers. The recombination of ions with their electrons primarily results in triplet-state excimers. Therefore, triplet states are much more prevalent in electron-recoil interactions than in nuclear-recoil interactions.

The enabling feature for PSD is that the triplet state has a longer decay time (the slow component of scintillation) compared to the singlet state (the fast component of scintillation). Therefore, with their longer triplet component, electron recoils produce more long-lived scintillation light than nuclear recoils. This difference in pulse shape can then be used to effectively discriminate neutron events from gamma-ray events, with observed rejection at the $10^{-6}$ to $10^{-8}$ level [1, 5, 10].

Several applications for liquid-argon detectors exist. A primary application is the attempted detection of weakly interacting massive particles (WIMPs), where a large active volume and powerful neutron-gamma and neutron-beta discrimination are required. The MiniCLEAN, DEAP-3600, ArDM, and DarkSide collaborations are building such detectors [11, 12, 13, 14]. LAr is also used for various neutrino measurements [12, 15, 16]. Others have used it as an active veto for neutrinoless double-beta-decay searches [1, 17].

Another application of LAr detectors is in active-interrogation measurements of cargo for security applications [7, 18]. The composition of the cargo material can be determined from the products released after interrogation of the cargo material with neutrons or other high-energy particles. The fast timing and PSD capability of LAr can be used to determine the identity and time structure of emitted products and characterize the interrogated material. Simple, cost-effective LAr detectors can be arrayed to cover a large area around the interrogated cargo to maximize detection efficiency. The fast decay time of the singlet state allows coincidence events to be identified, which can be important signatures of contraband nuclear materials [19].

Previous studies have shown that the addition to argon of a few parts per million of xenon results in improved PSD [1], better energy resolution [7], and faster decay time [20]. As show in Table I, pure argon produces a wavelength of 128 nm, with decay times of 7 ns and 1.6 μs. With the addition of a small amount of xenon, energy can be transferred to xenon excimers, which decay with the characteristic xenon decay times of 4.3 ns and 22 ns [21] and wavelength of 175 nm, thus making the bulk scintillation more xenon-like. In addition, a line at 149 nm appears due to decay of the lowest deep Xe state [22]. The combination of these two sets of decay times and the exchange of states between argon and xenon changes the time structure of the scintillation signal. With moderate dopant concentrations, the signal waveforms take on a characteristic two-hump shape (when plotted as a function of time) [23]. The early time hump decreases in amplitude as xenon-dopant concentration increases, while the late time hump grows and shifts to earlier times, eventually merging with the first [1, 23]. (This is the same evolution seen in Figure 4 below.) While the direct light yield is not expected to improve, previous studies have observed factor-of-two increases in detected light (signal yield), attributed to higher efficiency of the wavelength shifter at the longer wavelengths of xenon emission [20, 24]. This improved signal yield results in better energy resolution due to better statistics [1, 7, 20].

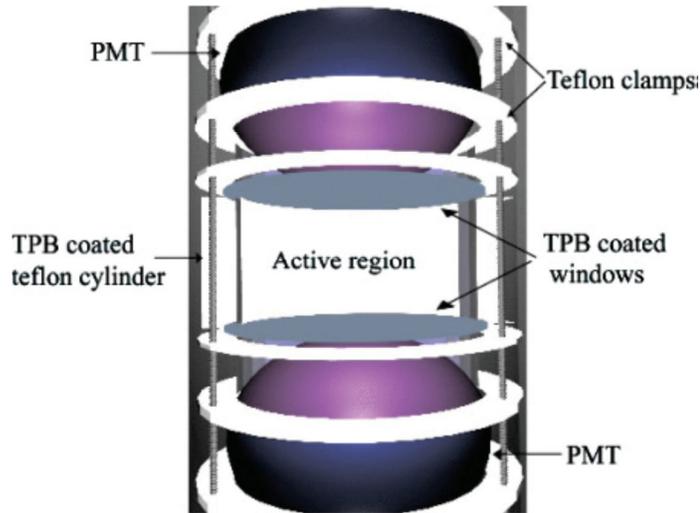

**Figure 1**. The internal structure of the microCLEAN detector used in this study.

As the xenon doping shifts much of the slow-light component to earlier times, the waveforms for different particles become more similar [1]. In itself, this will tend to reduce PSD, although the improved signal yield offsets this effect at first and results in slightly improved discrimination between alpha particles and gamma rays up to 300 ppm xenon [1].

In this work, we investigate PSD between neutrons and gamma rays in xenon-doped argon, which, to our knowledge, has not been reported elsewhere. Specifically, we report leakage of gamma-ray events into neutron-like pulses. We show the performance at a number of gamma-ray energies and a number of dopant concentrations. This study could influence the technical approaches toward the applications described above. In addition, by making a study of xenon doping with many steps of known quantities of xenon dopant, we replicate and extend previous observations of energy resolution, signal yield, and pulse shape.

After describing the experimental apparatus and standard analysis methods, the average waveform produced by gamma rays and by neutrons is presented as a function of dopant concentration. Fits are performed to estimate decay-time parameters, according to a mathematical model of the scintillation. Next, the signal yield and energy resolution are estimated from gamma-ray spectroscopy, and the neutron spectrum is used to infer the neutron energy scale. Pulse-shape discrimination is plotted as a function of energy for each dopant concentration. Finally, an observation of waveforms from interactions between the PMT and the wavelength-shifter-coated window is reported, which suggests the shape of the xenon-light-only waveform.

## 2. Experiment

We use the microCLEAN detector, described in other studies [5, 7, 25] and shown in Figure 1. It consists of a 10-cm-tall active region within a 20-cm-diameter Teflon cylinder. Windows of 3-mm-thick fused silica are on the top and the bottom of the active region, each in front of a curved Hamamatsu R5912-02-MOD 20-cm-diameter photomultiplier tube (PMT), 0.64-cm away from the window at closest approach. These tubes have a rise time of 4 ns and a transit time spread (FWHM) of 2.8 ns. Both the inside of the Teflon cylinder and the inner faces of the windows are coated with a thin layer of tetraphenyl butadiene (TPB) to shift the ultraviolet light to approximately 440 nm where the PMT is sensitive. Note that light from interactions in argon between the window and PMT will not be shifted before striking the nearby PMT. In addition, the fused-silica window is opaque to argon's 128-nm light, but somewhat transparent to xenon's 175-nm light. The glass PMTs are insensitive to both scintillation wavelengths. Therefore, light from argon that originates outside the active area will not be detected, but light from xenon that originates outside the active area may be (by traveling back through the fused silica window to the TPB layer).

This internal structure is placed in a 25-cm-diameter by 91-cm-tall stainless steel vessel into which liquid argon is dripped from a liquefier above. The argon vessel is located inside a vacuum Dewar with outer radius of 28 cm. The liquefier is cooled by a pulse-tube refrigerator and was run almost continuously for 380 days. During 270 days, including two months before data was recorded and the entire time that data was recorded, the argon was continuously pulled from the bottom of the volume, circulated through a hot gas-purification getter at about 1 SLPM, and returned through the liquefier. A custom slow-control program read pressures, temperatures, and flow rates, and controlled heater feedback to keep the temperature stable between 89 K and 92.5 K (absolute pressure between 1.3 and 1.6 bar). A post-experiment weighing of the full vessel gave $30 \pm 1$ kg as the total mass of argon in the vessel.

The PMTs were biased to either 1000 V or 1200 V, depending on the required gain, of approximately $7 \times 10^6$ or $5 \times 10^7$, respectively. Data were recorded from each PMT with a CAEN V1720 digitizer at 12 bits and 250 MS/s, and then transferred via a fiber-optic link to a PC for storage. Each collected waveform was 32.7 μs long, with the trigger occurring at 2.8 μs. Events could be read from the digitizer and stored at up to 2600 counts per second (cps). Figure 2 shows sample waveforms from the bottom PMT under two xenon-dopant concentrations.

For measurements of gamma rays, small sealed sources of a few μCi (several hundred kBq) were affixed to the surface of the outer Dewar, level with the center of the active region. For measurements of neutrons, a Thermo Electron MP320 deuterium-deuterium neutron generator – a source of 2.8-MeV neutrons – was placed in a borated polyethylene box and positioned level with the active region on another side of the detector.

Under normal operation, a trigger was generated whenever the signal from both the top and bottom PMTs were simultaneously beyond a threshold set to be about the amplitude of a single photoelectron pulse at 1200 V. Under this configuration, the trigger rate from background was about 700 counts per second (cps). To collect a clean gamma-ray sample, we used a $^{22}$Na source, decaying via positron emission into two 511-keV gamma rays emitted back to back. We used a 2-inch (5-cm) NaI(Tl) detector triggered in coincidence with the LAr detector to tag the paired gamma rays. The NaI(Tl) crystal surface was 18 cm radially out from the source. The NaI(Tl) signal passed through a 100x amplifier with 1-μs unipolar shaping. A trigger was sent to the DAQ when the rising edge of the trigger from both argon PMTs was within 1.1 μs of the time that the rising edge of a pulse from the NaI(Tl) detector passed 0.4 V (corresponding to about 100 keV). This coincidence requirement reduced the background trigger rate to below 1 cps. In post processing, further filters were applied, as detailed below.

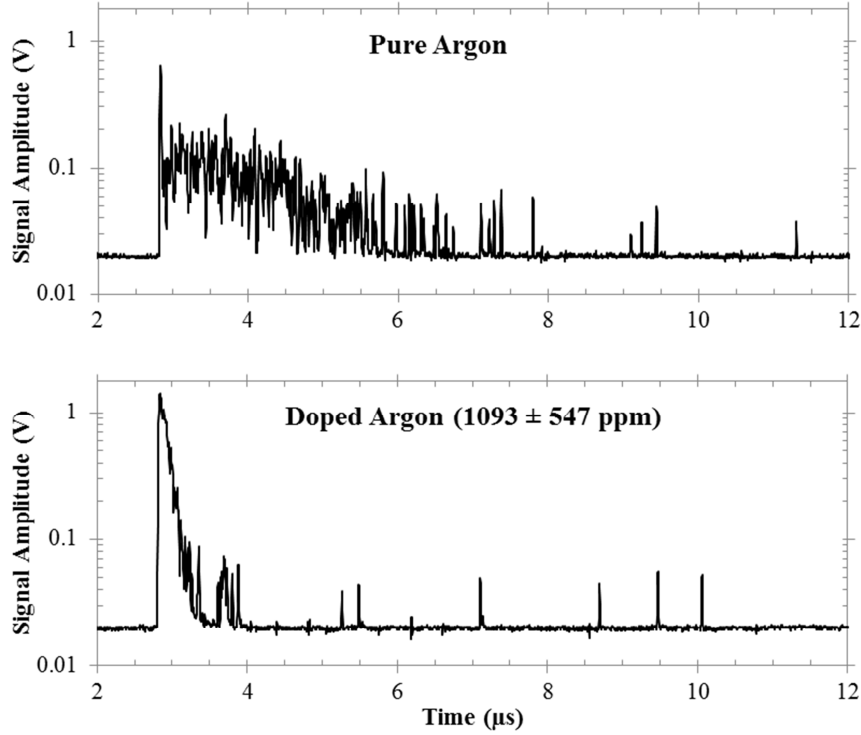

**Figure 2**. Representative waveforms from the bottom PMT biased to 1200 V, for a $^{57}$Co gamma-ray interaction. The upper waveform is from pure argon and the lower waveform is from doped argon at 1093 ± 547 ppm. In these examples, the rest of the waveforms beyond 12 µs are nearly flat. Both represent a similar number of collected photons.

When doping xenon into the argon volume, a small pipe segment (25.5 cm$^3$) or can (2072 cm$^3$) was evacuated, and then filled to a recorded pressure with xenon. This xenon sample was then opened to the argon circulation line and slowly allowed to diffuse into the detector to try to prevent freezing. Argon was then flushed through the can and allowed to mix before events were recorded. The concentration is reported in parts-per-million (ppm) by weight. Xenon is soluble in liquid argon to approximately 16% by weight at 87 K, and mixtures of up to 50,000 ppm xenon have been studied at this temperature without report of solubility problems [26, 27]. If xenon is frozen to the wall of a tube flowing pure argon gas, the xenon will sublimate into the argon with a flux given by the Hertz-Knudsen equation

$$J = \alpha P_{eq}(2\pi M_{Xe} K_B T)^{-1/2}$$

where $P_{eq}$ is the equilibrium vapor pressure of xenon, $M_{Xe}$ is the mass of a xenon atom, $K_B T$ is the thermal energy and $\alpha$ is a dimensionless constant near unity [28]. Solid xenon has an equilibrium vapor pressure of 1.7 Pa at the 83.8 K triple-point temperature of argon [29]. From this temperature, a lower bound of 4.3 × 10$^{22}$ / m$^2$s is calculated for the flux of frozen xenon sublimating from the walls of the tubing of our liquefier plumbing. This flux causes a solid xenon surface to recede at 3 × 10$^{-6}$ m/s, implying that the walls of a nearly plugged (but still flowing) tube should clear in about 800 seconds. Our system was allowed to circulate for at least 10 hours following each xenon addition to ensure complete mixing. The experiment was halted when xenon was introduced to the argon liquefier, where it froze and completely blocked the circulation path. This final addition was of similar size to to the prior addition, but was added into the main line more quickly. Prior to this addition, no evidence was observed of a partial blockage, nor of change in waveform shape from slow integration of frozen xenon into the argon.

## 3. Analysis

Analysis was performed using software modified from the UMImaging program described in [30]. UMImaging is a platform-independent standard C++ software library that is able to operate with a GUI or run with batch-jobs. It provides structures to read, store, and analyze data, with a focus on reconstruction methods. Once data files were collected, the analysis progressed by finding the gain of each PMT – from a Gaussian fit to a histogram of areas of single-photoelectron pulses in the waveform tails – then using these gains to calculate the total area of each waveform passing a set of analysis filters in terms of the number of single photoelectrons.

The total area of each waveform is calculated by first determining the baseline from the average of the first 600 samples (2.4 μs). A threshold is set very close to this baseline, much smaller than the height of a single photoelectron pulse, but above most electronic noise. The pulse area is then calculated as the sum of the area in each waveform region that contains at least one sample passing this threshold, with the region bounded by samples that return below the baseline.

### 3.1 Filters

Before events are included in other analyses, filters are applied to exclude poor events from further analysis.

The difference between the number of photoelectrons from the two PMTs must be less than 35% of the total, otherwise the event is eliminated. This asymmetry filter helps to remove events very near one PMT or the other.

Events with a waveform so large that it is within 0.1 V of the end of the digitizer range at any sample are also eliminated, to remove saturated events.

Pile-up events in which two or more events occur in the same waveform can contaminate the timing measurements. Because of the time structure of LAr/LXe scintillation (one large prompt pulse followed by many small delayed pulses), a basic cut to remove some of these events is one that removes any event where the first observed pulse in the waveform is not the largest pulse in the waveform. Although this cut clearly cannot remove pile-up events in which the second interaction was lower in energy than the first, it does remove approximately half while still preserving single-interaction events that consist of only a few photoelectrons.

The initial trigger time for each waveform is found from the time at which the waveform first reaches 10% of its maximum height for that event. Events where the trigger times of the two PMTs are more than 10 samples (40 ns) apart are thrown out. Additionally, these triggers must occur sooner than 9 μs from the end of the waveform to give sufficient length for the pulse-shape-discrimination integration window.

Finally, in data sets with a $^{22}$Na source and the NaI(Tl) channel, a filter is applied to only accept events in a 200-keV window around the 511-keV photopeak of the NaI(Tl). Because the $^{22}$Na source emits two back-to-back gamma rays at 511 keV, this filter requires that a 511-keV gamma ray was directed toward the center of the active region.

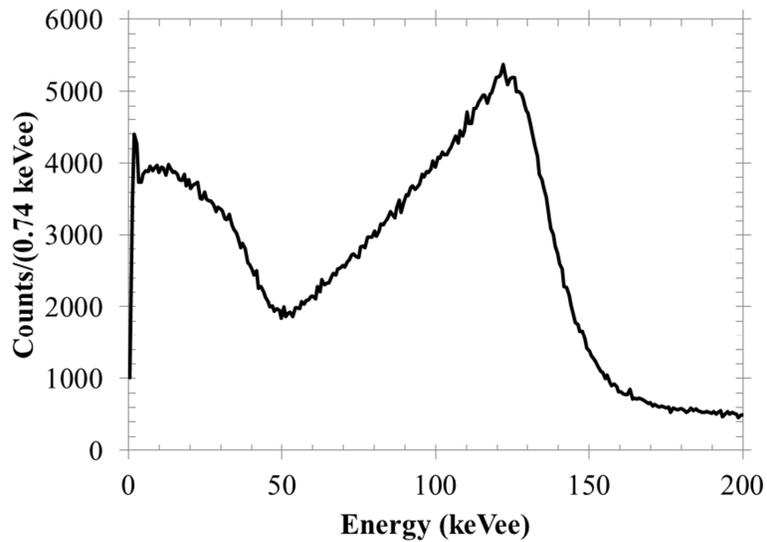

**Figure 3**. The $^{57}$Co spectrum from pure liquid argon.

## 4. Results and Discussion

### 4.1 Average Waveform Shape

At each dopant concentration, events were recorded from a $^{57}$Co sample and from a 2.8-MeV neutron source. Events in a range around the $^{57}$Co photopeak were used to create average waveforms; the range extended from the channel with the minimum counts in the Compton valley below the 122-keV peak, to the first channel above the peak with fewer counts than the valley. The $^{57}$Co spectrum for pure argon is shown in Figure 3. In this example, the range extended from 49 keVee to 145 keVee. The same range of photoelectrons was used for the neutron events at the same dopant concentration.

To further select neutron events, a prompt-fraction filter (see Section 4.6 for a description of the prompt-fraction discrimination method) was applied to the neutron data to remove, as much as possible, background gamma-ray events. For each dopant concentration, a histogram of the prompt fraction was generated for both neutron and gamma-ray events, and the neutron acceptance threshold was set in the valley between the two distributions.

For each PMT, the waveforms were aligned at the trigger time, interpolating between samples as needed, then averaged over at least $10^5$ events. The average waveforms from each PMT were then normalized by the single photoelectron size for each PMT and combined into a single average waveform. Figure 4 shows the resulting average waveforms for gamma rays and for neutrons at each dopant concentration. Because all come from a similar photoelectron cut, the photoelectrons per waveform are approximately the same for all waveforms, with higher areas corresponding to true signal-yield increases (see next section). The gamma-ray distributions are skewed to slightly larger yields relative to nuclear recoils due to the asymmetric shape of the $^{57}$Co peak, as shown in Figure 3.

**Figure 4**. (Next page) Average waveforms from neutrons and gamma rays in the photoelectron range around the $^{57}$Co photopeak, for each dopant concentration. All waveforms were aligned to time zero at 10% of their maximum amplitude. Solid lines are from gamma-ray waveforms and dashed lines are neutron waveforms. The left column uses logarithmic scales and the right column uses normal scales so all details can be observed.

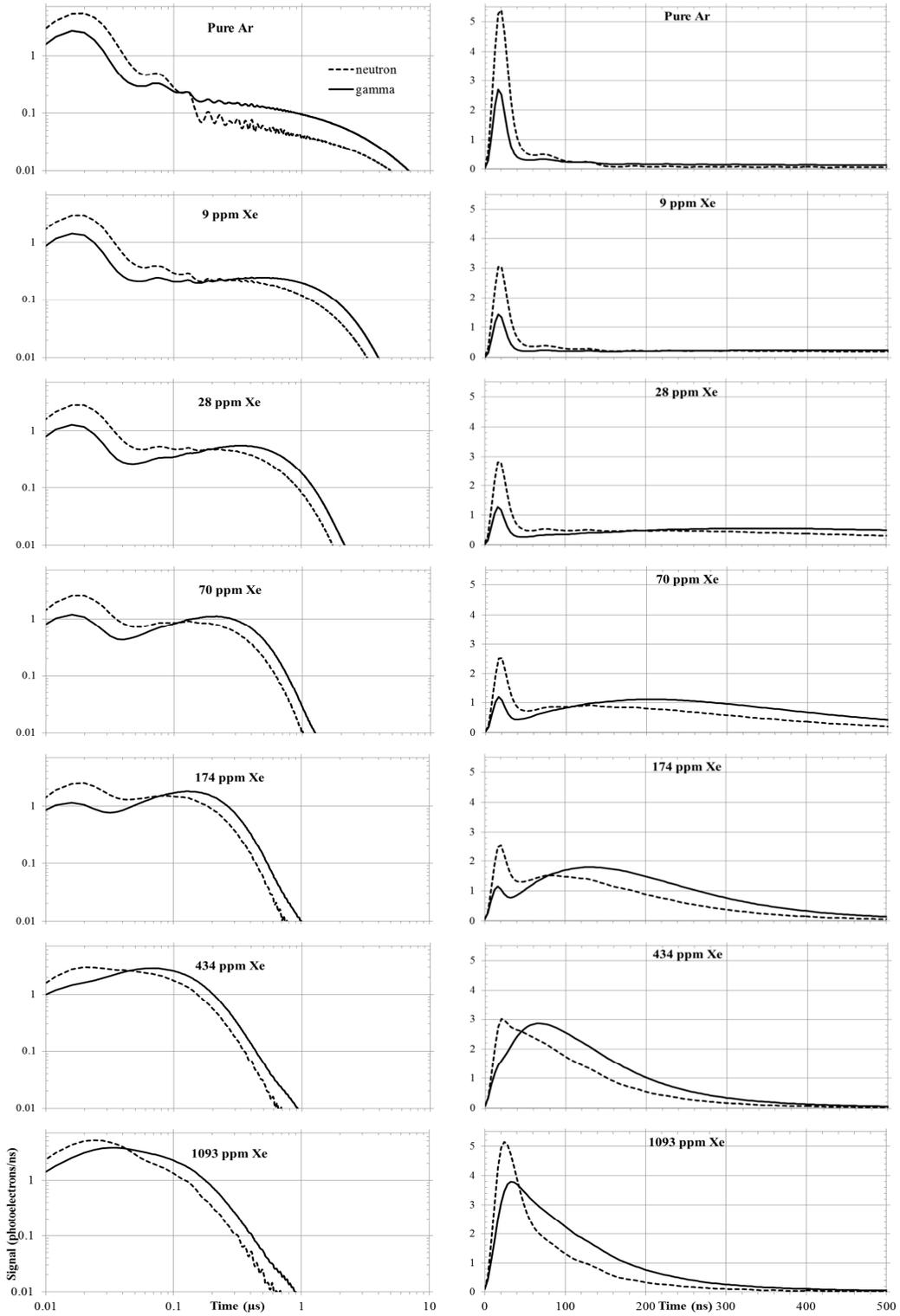

Several effects are obvious from Figure 4. With pure liquid argon, the waveform has a fast and a slow exponential-decay component, and the neutron-produced waveform has a larger fast component and smaller slow component, as expected. We also observe a "ringing," with a 56-ns period, on this and other waveforms that have a strong initial signal followed by a weak signal. This is attributed to incomplete termination effects on the signal cable (removing termination made the ringing amplitude larger). This problem could not be resolved during operation, as the bad connection was determined to be inside the inner vessel.

As xenon is doped into the argon, the decay time of the slow component reduces and a hump appears as argon excited states are transferred to the xenon, consistent with expectations. Though the decay time is similar between the gamma-ray- and neutron-produced waveforms, the time interval between the beginning of the waveform and the maximum of the hump is always shorter for neutron-produced waveforms. As doping continues, the second hump moves to earlier times while the first hump decreases in intensity. Eventually, at several-hundred ppm, the two humps merge.

## 4.2 Waveform Model

The shape of the waveforms can be modeled as described in [23]. In this model, an argon excimer is produced in either the singlet or triplet state. Singlet states decay quickly, but triplet-state excimers have enough time before decay to transfer energy to xenon excimers according to

$$Ar_2^* + Xe + \text{migration} \rightarrow (ArXe)^* + Ar$$
$$(ArXe)^* + Xe + \text{migration} \rightarrow Xe_2^* + Ar$$

This xenon excimer then decays with the characteristic decay times for the singlet and triplet states of xenon. The light production rate is the sum of the emission rates from the argon and xenon excimer emission rates. That is

$$r = \lambda_{Ar,1} N_{Ar,1} + \lambda_{Ar,3} N_{Ar,3} + \lambda_{Xe,1} N_{Xe,1} + \lambda_{Xe,3} N_{Xe,3}$$

where $N_{Ar,1}$ is the number of argon excimers in the singlet state and $N_{Ar,3}$ is the number of argon excimers in the triplet state, for instance, and the $\lambda$ are the decay constants of each state due to optical emission.

Let $\lambda_m$ be the rate at which the argon excimer converts to the combined argon-xenon excimer, $\lambda_d$ be the rate at which the argon-xenon excimer converts to the xenon excimer, $q$ be the fraction of argon excimers created in the singlet state, $p$ be the fraction of xenon excimers created in the singlet state, and $N_0$ be the initial number of argon excimers. In the original model [23], only triplet-state argon excimers could transfer to ArXe*. For completeness, we also include the transfer of singlet-state excimers to ArXe*. We assume the transfer rate is the same for both singlet and triplet transfers. The differential equations describing the transfer between states are then

$$\frac{dN_{Ar,1}}{dt} = -\lambda_{Ar,1} N_{Ar,1} - \lambda_m N_{Ar,1}$$
$$\frac{dN_{Ar,3}}{dt} = -\lambda_{Ar,3} N_{Ar,3} - \lambda_m N_{Ar,3}$$
$$\frac{dN_{ArXe}}{dt} = +\lambda_m N_{Ar,1} + \lambda_m N_{Ar,3} - \lambda_d N_{ArXe}$$
$$\frac{dN_{Xe,1}}{dt} = +\lambda_d p N_{ArXe} - \lambda_{Xe,1} N_{Xe,1}$$
$$\frac{dN_{Xe,3}}{dt} = +\lambda_d (1-p) N_{ArXe} - \lambda_{Xe,3} N_{Xe,3}$$

We define

$$\lambda_{1m} \equiv \lambda_{Ar,1} + \lambda_m$$
$$\lambda_{3m} \equiv \lambda_{Ar,3} + \lambda_m$$

Solving these equations, we find the rate of argon light output is

$$r_{Ar} = \lambda_{Ar,1} N_0 q\, e^{-\lambda_{1m}t} + \lambda_{Ar,3} N_0 (1-q) e^{-\lambda_{3m}t}$$

and the rate of xenon light output is

$$\begin{aligned}
r_{Xe} &= \frac{\lambda_{Xe,1}\lambda_d \lambda_m N_0 p\left((1-q)\lambda_{1m}+q\lambda_{3m}-\lambda_{Xe,1}\right)}{(\lambda_d-\lambda_{Xe,1})(\lambda_{Xe,1}-\lambda_{3m})(\lambda_{Xe,1}-\lambda_{1m})} e^{-\lambda_{Xe,1}t} \\
&+ \frac{\lambda_{Xe,3}\lambda_d \lambda_m N_0 (1-p)\left((1-q)\lambda_{1m}+q\lambda_{3m}-\lambda_{Xe,3}\right)}{(\lambda_d-\lambda_{Xe,3})(\lambda_{Xe,3}-\lambda_{3m})(\lambda_{Xe,3}-\lambda_{1m})} e^{-\lambda_{Xe,3}t} \\
&+ \frac{\lambda_m \lambda_d N_0 (p\lambda_d(\lambda_{Xe,3}-\lambda_{Xe,1})+\lambda_{Xe,3}(\lambda_{Xe,1}-\lambda_d))\left((1-q)\lambda_{1m}+q\lambda_{3m}-\lambda_d\right)}{(\lambda_d-\lambda_{3m})(\lambda_d-\lambda_{1m})(\lambda_{Xe,1}-\lambda_d)(\lambda_{Xe,3}-\lambda_d)} e^{-\lambda_d t} \\
&+ \frac{\lambda_d \lambda_m (p\lambda_{1m}(\lambda_{Xe,3}-\lambda_{Xe,1})+\lambda_{Xe,3}(\lambda_{Xe,1}-\lambda_{1m}))q N_0}{(\lambda_d-\lambda_{1m})(\lambda_{Xe,1}-\lambda_{1m})(\lambda_{Xe,3}-\lambda_{1m})} e^{-\lambda_{1m}t} \\
&+ \frac{\lambda_d \lambda_m (p\lambda_{3m}(\lambda_{Xe,3}-\lambda_{Xe,1})+\lambda_{Xe,3}(\lambda_{Xe,1}-\lambda_{3m}))(1-q) N_0}{(\lambda_d-\lambda_{3m})(\lambda_{Xe,1}-\lambda_{3m})(\lambda_{Xe,3}-\lambda_{3m})} e^{-\lambda_{3m}t}
\end{aligned}$$

Since the decay times for the xenon excimers are short, the exponential factors in the first two terms are almost zero. Therefore, the total light output rate is

$$r = A_1 e^{-\lambda_{1m}t} + A_2 e^{-\lambda_{3m}t} - A_3 e^{-\lambda_d t}$$

with constants $A_1$, $A_2$, and $A_3$ representing the remaining three terms.

Figure 5 shows the parameters resulting from the best fit of this equation to each waveform in Figure 4, where we define the fast decay time, the slow decay time, and the delay time as

$$T_f \equiv \frac{1}{\lambda_{1m}}$$
$$T_s \equiv \frac{1}{\lambda_{3m}}$$
$$T_d \equiv \frac{1}{\lambda_d}$$

Because there is a non-zero baseline in the waveforms, a positive constant baseline term was also added to $r$ as an additional parameter in the fit. The waveform fit was made in the time interval from just after the first peak in the waveform until the time at which the intensity fell to 0.01 photoelectrons/ns. In the few waveforms with strong "ringing", the ringing region was excluded from the fit. In most cases, the fit matched the data quite well, and changing the fit region slightly produced changes in the parameters of less than 10%. In the pure-argon case, the fits do not reproduce the step at the ringing near 100 ns.

As an additional test, the slow decay region was fit individually with simple exponentials plus a baseline. The slow component was fit to the region of the waveform with amplitude between 0.1 and 0.01 photoelectrons/ns (except for the waveform from pure argon with neutrons, where the region was shifted down by 30% because of the small amplitude of the neutron slow component in pure argon). The resulting parameters are the "limited fit" series in Figure 5b. Again, tests with varying the fit region gave less than 10% change in the fit parameters.

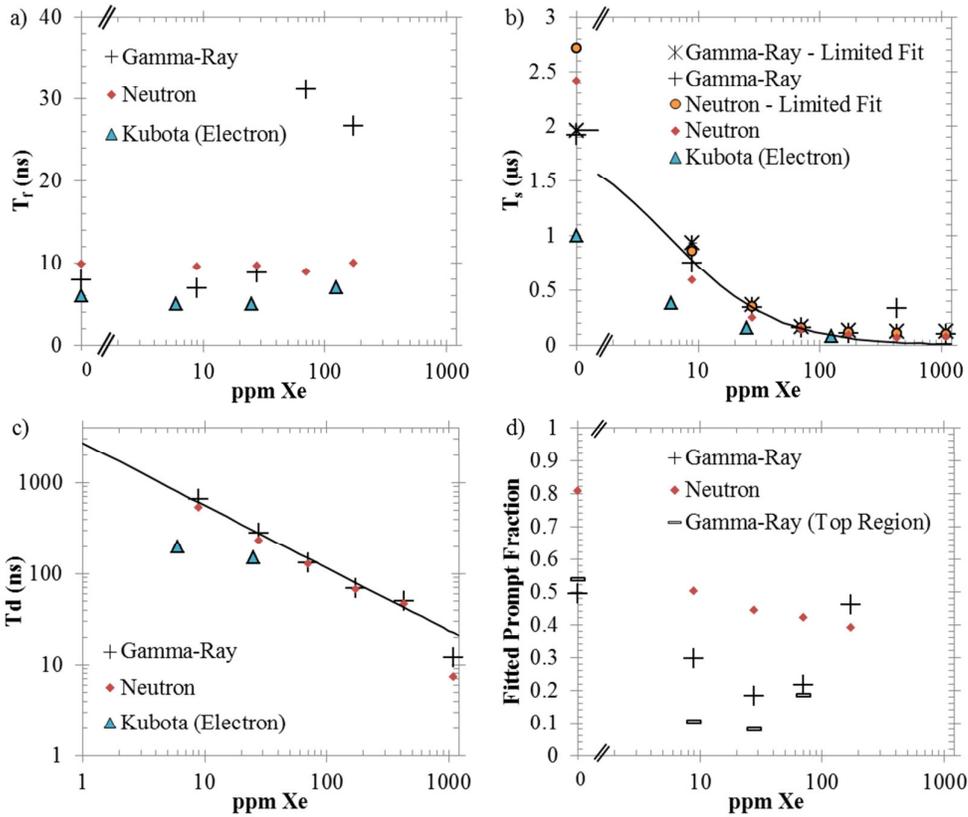

**Figure 5**. Fit parameters to the average waveforms in Figure 4 and a comparison with a previous study (triangles) [23]. For all parameter estimates, a minimum mean-squared error fit was performed from just after the first peak down to an intensity of 0.01 photoelectron/ns, eliminating regions with significant ringing. The "limited fit" in (b) is a fit to just the region with intensity between 0.1 and 0.01 photoelectrons/ns. The vertical error due to the choice of the fit region is about the size of the point in (a) through (c) and about 0.1 in (d). Horizontal error is the same as on other subsequent figures. The fit lines in (b) and (c) are described in the text. The "Gamma-Ray (Top Region)" series in (d) is from events thought to occur in the region between the PMT and the TPB-coated window and hence representative of pure xenon signals (as described in Section 4.7).

On all waveforms, the rise time (from first increase to reaching the peak) is about 20 ns, which may be due to the bandwidth of our signal-electronics chain and PMTs. Therefore, the decay time of the fast component may be systematically overestimated. Figure 5a shows that as the xenon concentration increases, the fast decay time stays steady at just below 10 ns for both neutrons and gamma rays. Compare this to the reported singlet decay time in argon of 7 ns. At large dopant concentrations after the humps in the waveform merge, it is hard to fit this decay time as there is no obvious first component. Therefore, the last two dopant concentrations (434 ppm and 1093 ppm) are not included in Figure 5a. The two abnormally high parameters in the gamma-ray waveform fits in the same figure for concentrations of 70 ppm and 174 ppm both give poor agreement with the waveforms in the fast-component region. However, fits to just that fast-component region, keeping other parameters constant, give decay times below 20 ns for the first point and 10 ns for the second point, more in agreement with the results from lower concentrations.

The slow decay time, in Figure 5b, drastically decreases as dopant concentration increases, explained by the transfer of argon excimers to argon-xenon excimers, according to the model. Since the transfer rate is assumed to be linear with xenon concentration, the plot of the slow decay time versus concentration in Figure 5b should be of the form $(\lambda_{Ar,3} + ac)^{-1}$ with constant $a$, pure-argon slow decay constant $\lambda_{Ar,3}$, and concentration $c$. This curve is traced on Figure 5b with $\lambda_{Ar,3} = 0.00051/\text{ns}$ and $c = 8.8 \times 10^{-5}/(\text{ppm} \cdot \text{ns})$. At high concentrations – above about 100 ppm – the $ac$ term is no longer linear with concentration and saturates with a minimum time constant $(ac)^{-1}$ of about 100 ns. According to the model, that term describes the time for an argon excimer to transfer to an argon-xenon excimer. As before, the abnormally high $T_s$ parameter in the gamma-ray fit at 434 ppm is due to a fit that does not capture the waveform shape well.

Figure 5c shows the transfer time constant from the argon-xenon to xenon excimer states. With pure argon, this component of the fit (that is, $A_3$) is very small, so the time constant is not plotted. For the others, very good agreement was found between the gamma-ray and neutron waveforms. As with the slow-decay-time parameter, we expect that the transition time should scale inversely with the concentration. However, the best fit to all but the last point, drawn in Figure 5c, follows $T_d = 2700c^{-0.7}$, where $c$ is the concentration in ppm and $T_d$ is in ns. In other words, the transition time does not decrease as much as one might expect due to an increase in the xenon concentration. Also, at the highest concentration, the transition time was measured to be a factor of two less than this.

Both neutron and gamma-ray data have similar fast and slow decay times, except with pure argon, in which the neutron slow decay time is significantly longer. Both data sets, with full and limited fitting regions, report longer slow-component decay times, especially with pure argon, than were found in previous studies [1, 5, 23, 31]. The cause of the difference is unclear. In addition, the argon-to-xenon transition time is longer in this study than has been reported by [23]. The longer transition time and slow decay time may have a common explanation. Nevertheless, the scintillation decay constants agree closely between the gamma-ray- and neutron-produced waveforms. Therefore, we conclude that the shape variation between neutron and gamma-ray waveforms is only due to the initial production quantity of each state rather than production of different states altogether. Figure 5d shows the ratio of the area of the fast component to that of the entire fitted waveform, $A_1 T_f / (A_1 T_f + A_2 T_s - A_3 T_d)$, for gamma rays and neutrons at each xenon concentration. It is clear that the greatest difference between the prompt fractions occurs with pure argon. The last anomalous gamma-ray point shown may be due to the poor fast-decay-region fit mentioned above. Again, at higher xenon concentrations, the fast component has merged with the second hump, so this quantity cannot be calculated there.

## 4.3 Signal Yield

The signal yield (detected-light yield) is found from the number of photoelectrons produced in both PMTs together at the 122-keV peak from $^{57}$Co and, separately, from the 662-keV peak from $^{137}$Cs or the 511-keV peak from $^{22}$Na. Past experiments with this detector system using pure argon reported signal yields of 4.85 and 6.0 photoelectrons per keV electron equivalent (keVee) [5, 32]. The current study found between 3.9 and 4.1 photoelectrons per keVee in pure argon. The decreased yield in this study may be due to degradation of the TPB wavelength shifter. Over the first three months of the experiment, with pure argon purified continuously through a getter, no significant change in the signal yield was observed. Therefore, we do not think the lower signal yield is due to impurities in the argon. In addition, our long scintillation decay time disfavors this hypothesis.

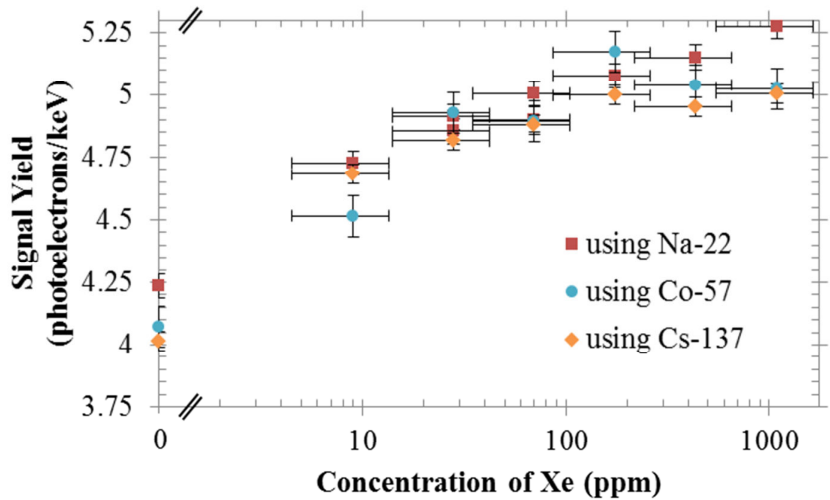

**Figure 6**. The signal yield from both PMTs at the 122-keV photopeak from a $^{57}$Co source, at the 511-keV photopeak from a $^{22}$Na source, and at the 662-keV photopeak from a $^{137}$Cs source, as a function of the amount of xenon dopant.

Figure 6 shows the signal yield as a function of xenon-dopant concentration. As dopant concentration increases, signal yield increases, consistent with previous studies that observed an increased light yield, explained by higher wavelength-shifter conversion efficiency. We do not believe this increase is due to time alone (due to reduction in impurity concentration, for instance), as the signal yield was stable at many measurements of pure argon, and then changed rapidly with the addition of the first xenon. At xenon concentrations of about 100 ppm, a possible plateau is observed in the signal yield, at approximately the same concentration as that at which the slow component decay time saturates, suggesting that the maximum fraction of excimer states are being transferred to xenon at this concentration. Finally, there is no systematic difference between the signal yields measured at the three different energies, from which we infer that scintillation efficiency is constant over the range of 122 keV to 662 keV.

**4.4 Energy Resolution**

With improved signal yield, even if the additional photons are correlated due to production in the wavelength shifter of multiple PMT-detectable photons for each original UV photon, we expect improved energy resolution. The spectrum from a number of different gamma-ray sources was recorded at each dopant concentration and the resolution for each is shown in Figure 7. A sample spectrum is shown in Figure 8 for $^{137}$Cs.

For these data sets, an additional filter is used to require that the difference in the number of photoelectrons from the two PMTs is less than 10% of the total number of photoelectrons. The energy resolution is found by first subtracting a linearly-interpolated continuum from under the peak, then, in the case of $^{137}$Cs and $^{22}$Na, fitting a Gaussian peak, or, in the case of $^{57}$Co, finding the full width at half max (FWHM) of the resulting peak. Variations due to choosing different fitting regions are reflected in the vertical error bars.

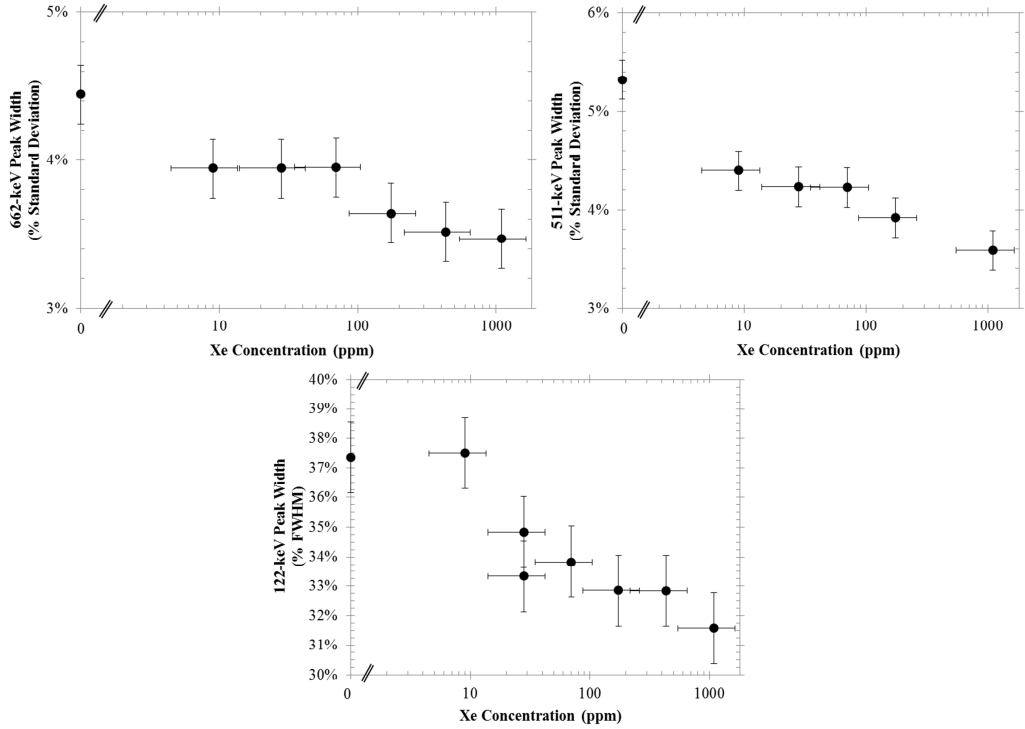

**Figure 7**. The resolution measured from the photopeak of a $^{137}$Cs source (662 keV), a $^{22}$Na source (511 keV), and a $^{57}$Co source (122 keV) as a function of xenon-dopant concentration for the center region of the detector. Note the $^{57}$Co resolution is presented in FWHM.

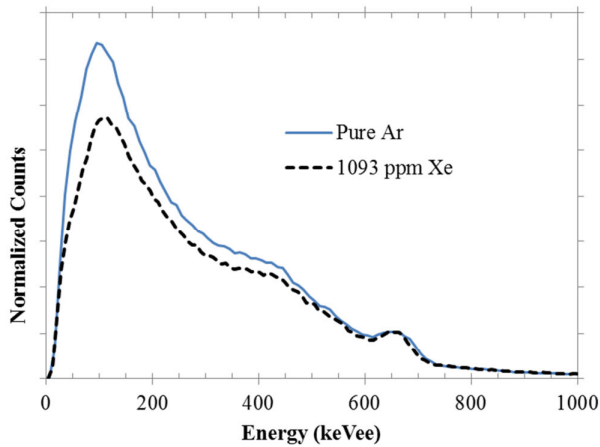

**Figure 8**. The spectrum from the center region of the detector from a $^{137}$Cs source, using pure argon and with doped argon. The spectra are normalized to the area in the center region of the photopeak at 662 keV.

We immediately see that the energy resolution improves as the dopant concentration increases, though the exact functional form of the improvement is unclear. This improvement is significantly more than would be expected just from the observed light-yield increase described above. For instance, using the resolution of 4.4% σ from the 662-keV peak in pure argon and the signal yield improvement, the resolution at 1093 ppm should only be 4.0%; however, it is actually 3.5%. The resolution improvement cannot be explained by the shorter length of the doped waveform causing less noise to be integrated into the pulse area; the baseline noise contributes much less than 0.1% to the overall resolution. Hence, doping of xenon appears to reduce fluctuation in detected photons. A previous study with this detector also observed resolution improvement to 3% from 5% σ at 662 keV when doping pure argon with about 2000 ppm xenon [7].

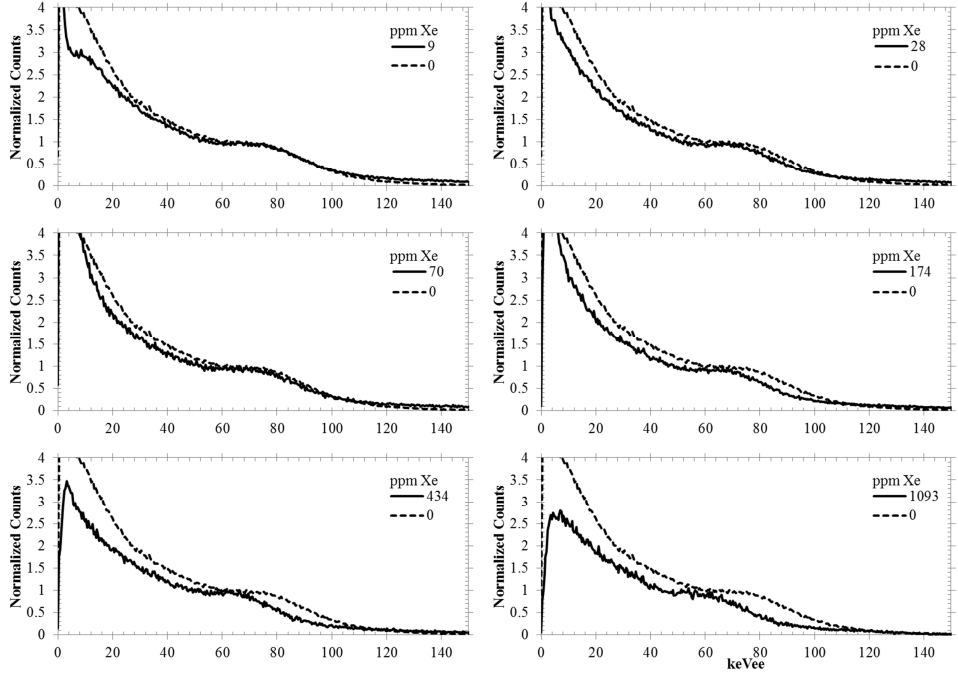

**Figure 9**. The neutron spectrum as a function of dopant concentration, normalized to the final plateau. The number of collected photoelectrons is converted to energy using the signal yield calculated for $^{57}$Co above. Note how the spectra appear to shift left at higher dopant concentrations.

### 4.5 Neutron Energy Scale

Before performing studies with neutrons, we must consider the neutron energy scale, which is known to be different from that of gamma-ray interactions due to reduction in the scintillation efficiency. The ratio between the light yield produced by nuclear recoils and electronic recoils of the same energy is known as $L_{eff}$. Furthermore, the scintillation yield may be non-linear with the energy of the particle, especially at low energies [4]. Figure 9 shows the spectrum recorded in our detector due to DD neutrons, using the energy scale obtained using the 122-keV peak from $^{57}$Co. Background gamma rays have been removed by requiring a minimum prompt fraction (see Section 4.6) in the waveforms. Again, this threshold is set in the valley between gamma-ray and neutron distributions in the prompt-fraction histogram. As the dopant concentration is increased, the neutron spectrum appears to shift to lower energy. In other words, the detectable-photon scintillation efficiency for neutrons is reduced and $L_{eff}$ reduces as xenon dopant increases. Roughly, the scintillation yield for neutrons relative to 122-keV gamma rays at the highest dopant concentration is 80% of that in pure argon. The absolute signal yield for neutrons still increases by nearly the same fraction as for gamma rays up to about 70 ppm xenon, but then reduces to almost that of pure argon by 1093 ppm xenon. More detailed analysis is difficult due to the relatively featureless shape of the neutron spectrum.

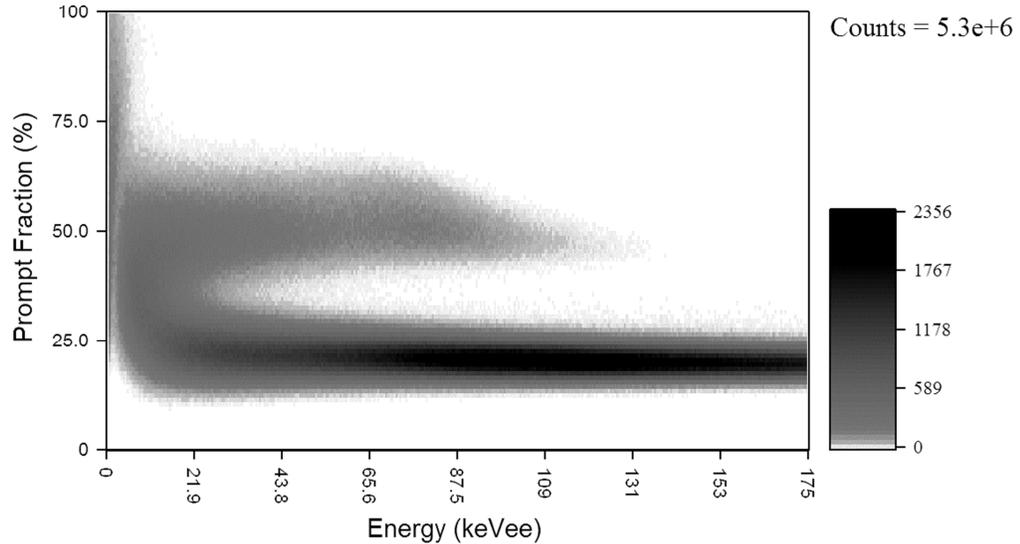

**Figure 10**. The prompt-fraction distribution (before 140 ns) as a function of energy for a DD neutron measurement in pure argon. The band with higher prompt fraction is identified as neutrons and the lower band is identified as gamma rays.

**4.6 Pulse-Shape Discrimination**

An important characteristic of LAr detectors is their very good discrimination between neutron and gamma-ray pulses based on the scintillation pulse shape. We use here a simple prompt-fraction pulse-shape discrimination statistic as defined in [5]

$$f_p = \frac{\int_{T_i}^{\xi} V(t)dt}{\int_{T_i}^{T_f} V(t)dt}$$

where $V(t)$ is the sum of the PMT waveforms, with the appropriate gains and baseline subtracted. Time zero is defined as the trigger time (at which the waveform reaches 10% of the maximum), $T_i$ is -50 ns to -10 ns, and $T_f = 9$ μs. Choosing other relatively large values of $T_f$ did not significantly change performance. Figure 10 shows an example distribution of prompt fraction, with $T_i = -10$ ns and $\xi = 140$ ns, as a function of energy from a measurement of neutrons in a gamma-ray background. The higher band, with prompt fraction near 50%, is due to neutrons, and the lower band, around 25% prompt, is due to gamma rays. In each energy bin, a histogram of prompt fraction is tallied and Gaussian functions fitted to the neutron and gamma-ray peaks. To achieve 50% neutron acceptance, the mean of the neutron peak is used as the acceptance threshold. Then, the electron recoil contamination (ERC) is the fraction of tagged gamma-ray events that have prompt fractions above this threshold. Gamma rays are tagged with a NaI(Tl) detector and $^{22}$Na source, as described above.

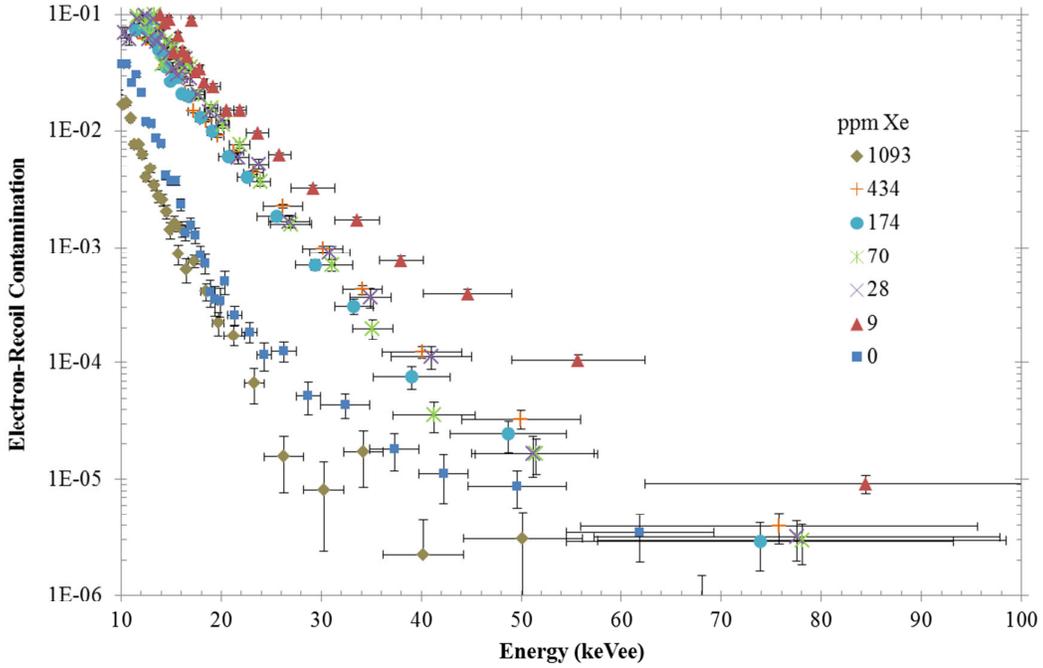

**Figure 11**. The fraction of tagged gamma-ray events leaking into the 50% neutron acceptance region of prompt fractions, as a function of deposited energy for each dopant concentration. Vertical error bars are statistical and horizontal error bars show the energy range integrated into the point.

The background-event count rate in the detector in coincidence with background events in the NaI(Tl) detector is 0.18% of the $^{22}$Na coincidence count rate, and from a prompt fraction distribution of background we estimate that fewer than 1% of background counts are due to neutrons. Therefore, with the $^{22}$Na count rate about four times greater than the background rate, background may limit the measurable ERC over all energies to somewhere below $7 \times 10^{-5}$.

Figure 11 shows the ERC as a function of energy using the optimal prompt-region parameter $\xi$ at each dopant concentration. The optimal prompt-region parameter $\xi$ was found by varying it over at least the range of 50 ns to 500 ns at each dopant concentration. With pure argon, very similar discrimination performance was achieved for a prompt-region parameter over a range from about 50 ns to over 200 ns (Figure 11 shows performance at $\xi = 150$ ns). With the first small addition of xenon, the discrimination performance shifts rapidly, with optimal performance occurring above 400 ns, extending to over 1000 ns (Figure 11 shows performance at $\xi = 500$ ns). As doping continues, this optimal range shifts back to shorter times, reaching 40 ns to 60 ns at 1093 ppm. In other words, the prompt region integral optimally covers the first hump until just before the peak of the second hump.

It is interesting to note that the PSD performance degrades with increasing xenon concentration until the second hump of the waveform merges with the first, when it rapidly improves. This is in contrast to [1], who reported improved discrimination with the first addition of xenon up to $300 \pm 80$ ppm, although this was between alpha particles and gamma rays, instead of between neutrons and gamma rays.

At first, even with improved signal yield from doped xenon, the presence of the second hump in the waveform degrades discrimination performance. After the first addition of xenon, however, performance slowly improves, possibly due in part to further light-yield improvement, but not completely explained by this. Between the highest two dopant concentrations, the prompt fraction in the neutron waveforms increases faster than that of the gamma-ray waveforms, improving the separation between their prompt fractions and improving discrimination performance.

**4.7 Waveforms Outside TPB Region**

One can make use of events that occur in the region between the PMT and the fused-silica window to determine the wavelength of light produced in each portion of the waveform. The fused-silica window is opaque to argon's 128-nm scintillation light, but at least somewhat transparent to xenon's 175-nm scintillation light. The PMT is insensitive to both wavelengths. Therefore, light from argon that originates in this area will not be detected, but scintillation light from xenon may enter the active region, interact with the TPB and produce a signal on the PMTs.

In fact, observing the distribution of gamma-ray events in prompt fraction and energy, an additional small-prompt-fraction band appears below that attributed to gamma rays. This band is only visible in data with xenon doping when removing the filters for similar number of photoelectrons on both PMTs and for pile-up in one waveform. These anomalous events have one PMT signal larger than the other, suggesting that the events are from regions very near the PMT faces. We believe that these are events interacting on the PMT side of the TPB-coated windows. The average waveforms for events in these regions are shown in Figure 12 and compared with the average waveforms from events in the center of the detector (as in Figure 4). Specifically, these are tagged gamma-ray events from the 511-keV peak in $^{22}$Na with an energy window of 30 keV to 130 keV (light equivalent). The center-region events are defined as having 40% to 55% of light on the top PMT and the top-region events are defined as having over 65% of light on the top PMT (strengthening this filter to 70% or more does not change the waveform shape significantly).

Events in the top region are almost absent in the pure argon case, representing only 0.2% of all events in this energy window. Because the PMT is insensitive to direct argon scintillation light and the fused-silica window is also opaque to argon scintillation light, we don't expect argon light in the region between the PMT and window to generate a PMT signal. Argon scintillation light is unable to reach the TPB wavelength shifter, which is only coated on the active-volume inner surfaces.

However, with xenon doping, the fraction of events in the top region increases to over 3.4%. Though the amplitude of the pulse on two PMTs differs, the shapes are very similar. We hypothesize that in this case, xenon scintillation light, with a longer wavelength, can penetrate the fused-silica window and reach the TPB wavelength shifter. Though the PMT is still insensitive to direct xenon light, it will see the secondary TPB light. Therefore, events emitting xenon scintillation light can be observed in the regions between the PMTs and silica windows. This is consistent with previous studies that have observed 175-nm light emission even with small concentrations of xenon dopant [22].

Figure 12 shows that the waveforms in the top region have a much-suppressed fast component and partially suppressed first portion of the slow component. According to the model in [23] described earlier, xenon light comes by way of the triplet state of argon and an intermediate ArXe* state. Kubota et al.'s model suggests that there should be no fast component coming from xenon, as xenon states must be preceded by an intermediate ArXe* state. The fact that some fast component is still observed suggests that some energy may be immediately transferred to the xenon. The small fraction of events included in this average that actually occur in the argon-light-sensitive active volume is too small to explain the fast-component hump in the absence of a fast-component in the true top-region events. Events with an additional Compton scatter in the active volume, which have energy below the low-energy threshold in pure argon, may contribute very slightly. The slow components in both regions can be explained by the model, as the first portion is due to the direct emission from argon triplet states and the last portion is due to the delayed xenon emission. This is consistent with the curves in Figure 12, in which the waveforms from both regions match in the last portions where the majority of light is from xenon.

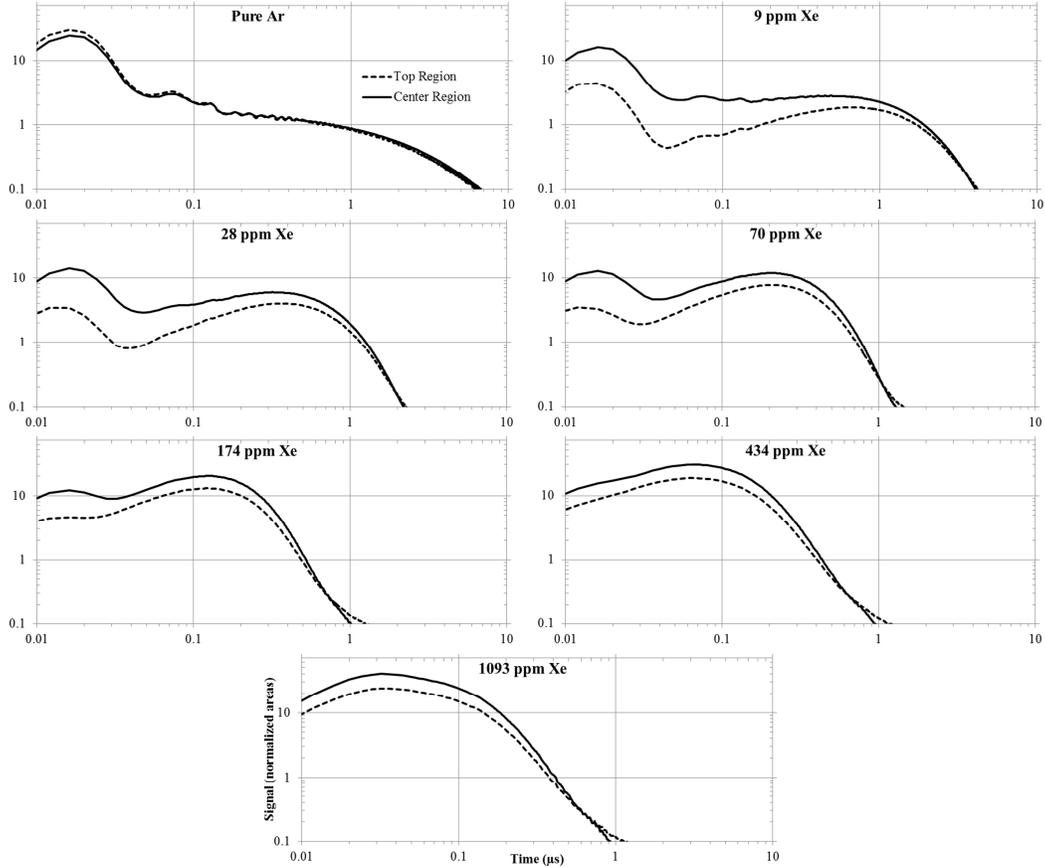

**Figure 12**. Average waveforms due to gamma-ray events in the center region of the detector and close to the top PMT, as a function of dopant concentration. The events near the top only record xenon light emission.

Fits were also performed for the average waveforms in the top region. Good agreement was found between the fit and data by keeping the same decay times as in Figure 5 and only changing the fraction of light from each component. Figure 5d shows the prompt fractions calculated with these new light fractions for the average waveforms in the top region. As expected, the prompt light is suppressed in the top-region average waveforms.

Without an absolute energy scale in the top region, it is hard to say what portion of energy is emitted as xenon light.

## 5. Conclusions

This study investigated the scintillation response of xenon-doped liquid argon with six different xenon concentrations. Qualitatively, the waveform shape matched that of previous studies [23], with a second hump appearing after doping, which moved to shorter times with increased dopant concentration. According to the migration model described above, this hump is the consequence of argon excimers transferring to xenon excimers. Since xenon excimer decay time, in both the singlet and triplet states, is shorter than the decay time of the argon triplet state, the competing interaction transferring energy to xenon shifts the observed slow decay time faster. The extent of this shift is proportional to the xenon concentration, as the individual transfers are more probable when more xenon is nearby. We observed this, but with a saturation of the transfer rate at a time constant of about 100 ns. The faster total decay time of doped argon can greatly improve its usefulness for applications such as active-interrogation, where a high count rate is essential.

According to the model, the first hump in the waveform is due to the singlet state of argon, and the second hump is due to both the argon triplet and xenon de-excitation. When observing the waveforms from regions of the detector that collect only xenon scintillation light, we saw the expected suppression of the first hump and a portion of the second hump. However, the first hump was still present, suggesting that some xenon scintillation light is produced much more quickly than is expected from the model.

By fitting the model of scintillation light production to both neutron- and gamma-ray-produced waveforms, we found very similar decay-time parameters for all cases, confirming that the same states are being created in each case. The variation in waveform shape between the two types of waveforms is due to relative quantity of each state.

The relative intensity of the fast and slow components of the waveform has been used to good effect in PSD measurements. The immediate effect of adding xenon dopant is to shift some of the slow component into the prompt window. In the absence of higher light or signal yield, which independently improves discrimination, discrimination based on prompt fraction should get worse because the fast and slow components overlap more in time. However, we have observed that after the initial degradation in PSD, PSD improved at the xenon concentrations where the second hump began to merge with the first. Improved PSD can be very important for measurements, such as WIMP searches, where strong discrimination is essential.

Energy resolution is important in some applications. The addition of xenon dopant was found to improve energy resolution to $(3.5 \pm 0.2)\%$ $\sigma$ at 662 keV, compared with a pure-argon resolution of $(4.4 \pm 0.2)\%$ $\sigma$. Some but not all of this improvement may be due to the improved signal yield – the shift from $4.0 \pm 0.1$ photoelectrons/keV to $5.0 \pm 0.1$ photoelectrons/keV with dopant. Very little of this improvement can be explained by the shorter waveform length, including less noise.

Finally, in this study, we observed an apparent downward shift in the neutron spectrum relative to the gamma-ray energy scale, as dopant concentration increased. This suggests there may be additional gain or non-linearity introduced by xenon doping, which must be considered in future applications.

## Acknowledgments

This work was supported in part by Adelphi Technology, LLC via NAVEODTECHDIV / Engility Corporation Contract No.: N00178-04-D-4143-FG01, Subcontract 10-110, Program Element Holder: Physical Security Equipment Action Group (PSEAG) / RDT&E#: 0603161D8Z. The authors wish to thank Zhong He at the University of Michigan for use of the base for the software used for analysis [30].